\documentclass[10pt,journal,compsoc]{IEEEtran}
\usepackage{ifpdf}
\usepackage{float}
\usepackage{graphicx}
\usepackage{multirow}
\usepackage{csquotes}
\usepackage{diagbox} 
\usepackage{enumerate}
\usepackage{algorithm}
\usepackage{flushend}
\usepackage{enumitem}
\usepackage{ragged2e}
\usepackage{hyperref}
\usepackage{array}
\usepackage{hyperref}
\usepackage{amssymb}
\usepackage{makecell}
\usepackage{color}
\usepackage{cite}
\usepackage{enumitem}
\usepackage{color,soul}
\usepackage{pbox}
\usepackage{amssymb}

\ifCLASSINFOpdf
 \else
 \fi
\usepackage{amsmath}
\usepackage{amsfonts}
\usepackage{algorithmic}

\usepackage{url}

\usepackage{nomencl}
\makenomenclature

\makeatletter
\renewcommand\subsection{\@startsection{subsection}{2}{\z@}%
                                     {-3.25ex\@plus -1ex \@minus -.2ex}%
                                     {-1.5ex \@plus .2ex}%
                                     {\normalfont\large\bfseries}}
\renewcommand\subsubsection{\@startsection{subsubsection}{3}{\z@}%
                                     {-1.00ex\@plus -1ex \@minus -.2ex}%
                                     {-0.1ex \@plus .2ex}%
                                 {\normalfont\normalsize\itshape}}
\makeatother

\begin{document}

\title{BIDEAL: A Toolbox for Bicluster Analysis - Generation, Visualization and Validation}

\author{Nishchal K.~Verma,~
        T.~Sharma,~
        S.~Dixit,~
        P.~Agrawal,~
        S.~Sengupta,~
        and~V.~Singh
\IEEEcompsocitemizethanks{\IEEEcompsocthanksitem Nishchal K. Verma, T. Sharma, S. Dixit, P. Agrawal, and V. Singh are with the Dept. of Electrical Engineering, IIT Kanpur, India. S. Sengupta is with Dept. of Electrical Engineering, Jadavpur University, India. \protect 
E-mail: nishchal@iitk.ac.in.}
\thanks{The developed software is available for academic and research use at \url{http://www.iitk.ac.in/idea/bideal/}.}
}

\IEEEtitleabstractindextext{

\begin{abstract}
\justifying
This paper introduces a novel toolbox named  BIDEAL for the generation of biclusters, their analysis, visualization, and validation. The objective is to facilitate researchers to use forefront biclustering algorithms embedded on a single platform. A single toolbox comprising various biclustering algorithms play a vital role to extract meaningful patterns from the data for detecting diseases, biomarkers, gene-drug association, etc. BIDEAL consists of seventeen biclustering algorithms, three biclusters visualization techniques, and six validation indices. The toolbox can analyze several types of data, including biological data through a graphical user interface. It also facilitates data preprocessing techniques i.e., binarization, discretization, normalization, elimination of null and missing values. The effectiveness of the developed toolbox has been presented through testing and validations on Saccharomyces cerevisiae cell cycle, Leukemia cancer, Mammary tissue profile, and Ligand screen in B-cells datasets. The biclusters of these datasets have been generated using BIDEAL and evaluated in terms of coherency, differential co-expression ranking, and similarity measure. The visualization of generated biclusters has also been provided through a heat map and gene plot.
\end{abstract}

}

\maketitle
\nomenclature{$a_{IJ}$}{Mean of all elements}
\nomenclature{$a_{Ij}$}{Mean of the $j^{\text{th}}$ gene}
\nomenclature{$a_{iJ}$}{Mean of the $i^{\text{th}}$ gene}
\nomenclature{$a_{ij}$}{Expression level of instance $i$ under attribute $j$}
\nomenclature{$\textbf{A}$}{Data matrix}
\nomenclature{$\textbf{B}_i$}{$i^{\text{th}}$ Bicluster}
\nomenclature{$B_{num}$}{Number of biclusters}
\nomenclature{$C$}{Set of vertices indicating columns of matrix}
\nomenclature{$E$}{Set of edges of graph}
\nomenclature{$F(\cdot)$}{Indicator function}
\nomenclature{$G$}{Bipartite graph}
\nomenclature{$I$}{Subset of rows/ genes}
\nomenclature{$\mid I \mid$}{Number of genes}
\nomenclature{$I_r$}{Row sets of the $r^{\text{th}}$ sub-matrix}
\nomenclature{$J$}{Subset of columns/ conditions}
\nomenclature{$jac$}{Jaccard index function}
\nomenclature{$\mid J \mid$}{Number of conditions}
\nomenclature{$jac$}{Jaccard index function}
\nomenclature{$J_r$}{Column sets of the $r^{\text{th}}$ sub-matrix}
\nomenclature{$\kappa_{jk}$}{Membership indicator variable of $j^{\text{th}}$ column of $k^{\text{th}}$ bicluster}
\nomenclature{$l_r$}{Level of $r^{\text{th}}$ sub-matrix}
\nomenclature{$\varphi(\textbf{D})$}{Significant score of  $\textbf{D}$ sub-matrix}
\nomenclature{$O$}{Optimal set of vertices}
\nomenclature{$Q$}{Measure for B-type co-expression}
\nomenclature{$R$}{Set of vertices indicating rows of matrix}
\nomenclature{$\rho_{ik}$}{Membership indicator variable of $i^{\text{th}}$ row of $k^{\text{th}}$ bicluster}
\nomenclature{$\theta_{ijk}$}{Background layer of element in $i^{\text{th}}$ row of $j^{\text{th}}$ column of $k^{\text{th}}$ bicluster}
\nomenclature{$T$}{Measure for T-type co-expression}
\nomenclature{$\textbf{u}$}{Prototype column vector}
\nomenclature{$\textbf{v}$}{Vector of factors}
\nomenclature{$\xi$}{Additive noise $\in N(0,1)$ where $N$ is the Normal distribution}
\printnomenclature

\section{Introduction}
\label{sec:Introduction}
\IEEEPARstart{B}{iclustering} has become prevalent and useful data mining technique among researchers for analyzing the data. It has been applied to a wide variety of applications such as bioinformatics, information retrieval, text mining, dimensionality reduction, recommender systems, electoral data analysis, disease identification, association rule discovery in databases, and many more \cite{ref:madeira}. Among these, bioinformatics \cite{ref:nano} \cite{ref:TCBB} seems to have taken the advantage of biclustering for analysis of the gene expression data. During any biological process under different experimental conditions, genes are examined by their expression levels. The data is present in a matrix form with rows representing genes and columns as experimental conditions. The aim is to group genes and conditions into a sub-matrix to obtain crucial biological information such as identification of co-regulated patterns among genes. A bicluster $\textbf{B}$ can be represented as
\begin{gather}
\textbf{B}=\begin{bmatrix}
    b_{11} & b_{12} & b_{13} & \dots  & b_{1|J|} \\
    b_{21} & b_{22} & b_{23} & \dots  & b_{2|J|} \\
    \vdots & \vdots & \vdots & \ddots & \vdots \\
    b_{|I|1} & b_{|I|2} & b_{|I|3} & \dots  & b_{|I||J|}
\end{bmatrix}
\end{gather}
where $b_{ij}$ refers to the expression level of instance $i$ under sample $j$, $\forall ~i \in \{ 1, 2, ..., |I| \}$ and $\forall ~j \in \{ 1, 2, ..., |J| \}$, $|I|$ is the number of instances, and $|J|$ is the number of attributes. It involves finding the maximum sub-matrices in a data matrix with maximum coherency. Since biclustering is a NP-hard problem, various heuristics and meta-heuristics approaches have been used in the literature to find better solutions \cite{ref:review}.

The traditional clustering algorithms give equal importance to all the columns. These algorithms are $K$-means clustering \cite{ref:Mac}, hierarchical clustering \cite{ref:Jhonson}, self-optimal clustering \cite{ref:NKOPtimal}, improved mountain clustering \cite{ref:NKIMC}, fuzzy C-means clustering \cite{ref:Bezdek}, unsupervised fuzzy clustering\cite{ref:ufc}, etc. Each algorithm has its own advantage. Despite their usefulness, they are not very helpful in a variety of problems. For example, every gene may not take part in every condition with gene expression analysis. Thus, combinatorial regulation and joint patterns of gene expression biclustering are essential to realize the complex nature of genes. In \cite{ref:NKbiclusterCompare}, a plethora of solutions to perform biclustering has been presented. Undoubtedly, among the pool of algorithms, all have their own distinctive ways including heuristic and statistical approaches with their merits and demerits. It is not expected that a single approach would turn out to be well-suited for all types of data. So, any problem must be tackled with respective suitable algorithms and the best result must be noted. This generates the need of a comprehensive biclustering toolbox where various algorithms can be tested, validated, and visualized. A toolbox can be compared in terms of the following:
\begin{enumerate}[label=(\alph*)]
    \item Number of algorithms embedded in the toolbox.
    \item Number of validation indices present for qualitative analysis of generated biclusters.
    \item Number of visualization methods available for generated biclusters.
    \item User-friendly interface of the toolbox.
\end{enumerate}
\begin{table*}
\caption{Summary of the biclustering toolboxes}
\label{table:Summary of the Previous Biclustering Toolboxes}
\centering
\resizebox{.98\textwidth}{!}{%
\begin{tabular}{|c|c|c|c|c|}
  \hline 
  \textbf{Toolboxes} & \textbf{Algorithms} & \textbf{Validation Indices} & \textbf{Visualization Methods} & \textbf{Platform} \\
  \hline 
  \textbf{BicAT}\cite{ref:BicAT} & CC\cite{ref:CC}, ISA\cite{ref:isa}, OPSM\cite{ref:OPSM}, xMotif\cite{ref:xMotif} & None & Heat Map\cite{ref:ClusterHeat} & JAVA\\
  \hline
  \textbf{BiVisu}\cite{ref:BiVisu} & \makecell{Greedy version of pCluster} & \makecell{Mean Square Residue,\\ Average Correlation Value} & \makecell{Heat Map, Parallel\\ Coordinate Plots} & MATLAB\\
  \hline
  \textbf{BicOver-lapper 2.0}\cite{ref:BicOverlapper} & Visualization Toolbox & None & Venn like Diagrams & R, JAVA\\
  \hline
  \textbf{Expander}\cite{ref:Expander} & SAMBA\cite{ref:SAMBA} & None & Heat Map & JAVA\\
  \hline
  \textbf{BAT}\cite{ref:BAT}& BiHEA\cite{ref:Bihea} & Pairwise Gene Analysis & Heat Map, Numerical Matrix & JAVA\\
  \hline
  \textbf{BiBench}\cite{ref:BiBench} & \makecell{CC, OPSM, xMotif, kSpectral\cite{ref:kspectral}, ISA, Plaid\cite{ref:plaid}, BiMax\cite{ref:BiMax},\\ Bayesian, QUBIC\cite{ref:QUBIC}, FABIA\cite{ref:FABIA}, COALESCE} & Jaccard Index\cite{ref:Fili}, F-measure & \makecell{Heat Map, Bicluster\\ Projection, Parallel Coordinates} & Python\\
  \hline
  \textbf{BiClust}\cite{ref:biclust} & BiMax, CC, Plaid & Jaccard Index, Constant Variance & \makecell{Parallel Plot, Heat\\ Map, Bubble Plot} & R\\
  \hline
  \textbf{BicNET}\cite{ref:BicNET} & BicNET & None & Biclustering Network Data & Java\\
  \hline
  \textbf{MTBA}\cite{ref:MTBA} & \makecell{CC, BSGP\cite{ref:BSGP}, ISA, OPSM, kSpectral, ITL\cite{ref:ITL}, xMotif, BiMax,\\ Plaid, FLOC\cite{ref:Floc}, BiMax, LAS\cite{ref:LAS}} & \makecell{Jaccard Index, SB Score,\\ Constant and Sign Variance} & Heat Map, Gene Plot & MATLAB\\
  \hline
  \textbf{CoClust}\cite{ref:Coclust} & \makecell{Modularity Based, Information-Theoretic Based} & None & \makecell{Cluster Plot, Cluster Size,\\Heat Map, Cluster Graph} & Python\\
  \hline
  \textbf{BicPAMS}\cite{ref:Bicpam} & \makecell{BicPAM, BicNET, Bic2PAM, BiP, BiModule} & None & \makecell{Graphical Display,\\Heat Map} & Java\\ 
  \hline
  \makecell{\textbf{BIDEAL}\\ \textbf{(Proposed Toolbox)}} & \makecell{CC, BSGP, OPSM, ISA, kSpectral, ITL, xMotif, Plaid, FLOC, BiMax,\\ LAS, FABIA, BitBit\cite{ref:BitBit}, BiSim\cite{ref:BiSim}, MSVD\cite{ref:MSVD}, QUBIC, ROBA\cite{ref:ROBA}} & \makecell{Jaccard Index, SB Score, Constant and\\ Sign Variance, Hausdorff, MSE} & \makecell{Heat Map, Gene Plot, Cluster\\ Plot, Numerical Matrix} & MATLAB\\
  \hline
\end{tabular}
}
\end{table*}

Based on the above-mentioned features, it can be summarized that a toolbox must be diverse in nature. In the past decade, the growing demand of biclustering algorithms has led the intense research on developing toolboxes for biclustering. This paper proposes a user-friendly toolbox namely \enquote{BIDEAL} which incorporates $17$ biclustering algorithms, $6$ validation indices, and $3$ visualization methods. Table \ref{table:Summary of the Previous Biclustering Toolboxes} summarizes various biclustering toolboxes in terms of available algorithms, validity indices, and visualization methods. Considering the visualization methods or result presentation for generated biclusters, BicAT \cite{ref:BicAT}, BicOver-lapper 2.0 \cite{ref:BicOverlapper}, Expander \cite{ref:Expander}, and BicNET \cite{ref:BicNET} provide only single visualization method. On the other hand, BiVisu \cite{ref:BiVisu}, BAT \cite{ref:BAT}, BiBench \cite{ref:BiBench}, BiClust \cite{ref:biclust}, MTBA \cite{ref:MTBA}, CoClust \cite{ref:Coclust}, BicPAMS \cite{ref:Bicpam}, and BIDEAL have multiple methods of visualization. Among these, CoClust and BIDEAL offers the maximum number of visualization methods. By default, BIDEAL provides bicluster results in a numerical matrix. Another important feature of a toolbox is the validation indices to check the quality of obtained biclusters. BiVisu, BAT, BiBench, and BiClust offers only one or two validation indices whereas, BIDEAL have six i.e. maximum among the listed toolboxes. The Graphical User Interface (GUI) of any application for the execution of various algorithms on a single platform alleviates the process. The user-friendly interface of BIDEAL enables the testing of new dataset quite easy without any prior knowledge of back-end programming. On the other hand, BiBench, BiVisu, BiClust, CoClust, and MTBA requires a little bit familiarity with the programming knowledge. Moreover, BicAT allows the execution of algorithms with default parameter settings, which is a constraint whereas, BIDEAL allows to change these parameters.

\textit{Contributions:} This paper introduces the proposed BIDEAL toolbox, its necessity, and importance in comparison with other existing toolboxes in literature. Table \ref{table:Comparison of features comprised with different biclustering toolboxes} summarizes the comparison of the features available in BIDEAL with respect to existing toolboxes in the literature. In summary, the features of BIDEAL are as follows:
\begin{enumerate}[label=(\alph*)]
    \item It is developed to integrate the largest number of biclustering algorithms, validation indices, and visualization methods (over existing toolboxes) on a single platform.
    \item It accommodates preprocessing methods as well within.
    \item It has a user-friendly interface than other existing biclustering toolboxes.
    \item To demonstrate the usefulness of BIDEAL, it has experimented with four standard datasets and their validation indices have been compared.
\end{enumerate}
To the best of our knowledge, no existing biclustering toolboxes have all these features incorporated on a single platform.

The paper is arranged as: Section \ref{sec:BIDEAL: Biclustering Algorithms} presents a brief introduction about biclustering algorithms embedded in BIDEAL, Section \ref{sec:Validation Indices} describes validation indices, Section \ref{sec:BIDEAL: Key Features and GUI} illustrates GUI of BIDEAL, and Section \ref{sec:Case Studies} provides the results on four standard datasets using BIDEAL. Finally, Section \ref{sec:Conclusions} concludes the paper.

\begin{table*}[ht]
  \caption{Comparison of the features comprised with various biclustering toolboxes}
  \label{table:Comparison of features comprised with different biclustering toolboxes}
  \centering
  \resizebox{.98\textwidth}{!}{%
  \begin{tabular}{|c|c|c|c|c|c|c|c|c|c|c|c|c|}
  \hline 
    \backslashbox{\textbf{Features}}{\textbf{Toolboxes}} & \makecell{\textbf{BicAT}\\ \cite{ref:BicAT}} & \makecell{\textbf{BiVisu}\\ \cite{ref:BiVisu}} & \makecell{\textbf{BicOver-lapper}\\ \textbf{2.0} \cite{ref:BicOverlapper}} & \makecell{\textbf{Expander}\\ \cite{ref:Expander}} & \makecell{\textbf{BAT}\\ \cite{ref:BAT}} & \makecell{\textbf{BiBench}\\ \cite{ref:BiBench}} & \makecell{\textbf{BiClust}\\ \cite{ref:biclust}} & \makecell{\textbf{BicNET}\\ \cite{ref:BicNET}} & \makecell{\textbf{MTBA}\\ \cite{ref:MTBA}} & \makecell{\textbf{CoClust}\\ \cite{ref:Coclust}} & \makecell{\textbf{BicPAMS}\\ \cite{ref:Bicpam}} & \makecell{\textbf{BIDEAL} \\\textbf{(Proposed Toolbox)}} \\
  \hline
  \textbf{No. of Algorithms} & 5 & 1 & 1 & 1 & 1 & 4 & 3 & 1 & 12 & 3 & 5 & \textbf{17} \\
  \hline
  \textbf{\makecell{No. of Validation Indices}} & 0 & 2 & 0 & 0 & 1 & 2 & 2 & 0 & 4& 0 & 0 & \textbf{6} \\
  \hline
  \textbf{\makecell{No. of Visualization Methods}} & 1 & 2 & 1 & 1 & 3 & 3 & 3 & 2 & 2 & 4 &2 & \textbf{5} \\
  \hline
  \textbf{\makecell{Graphical User Interface (GUI)}} & Yes & No & Yes & Yes & Yes & No & No & Yes & No & No & Yes & \textbf{Yes} \\
  \hline
  \end{tabular}
  }
  \textit{The values shown in bold represents the best feature among all the toolboxes.}
\end{table*}

\section{BIDEAL: Ready for use Biclustering Algorithms}
\label{sec:BIDEAL: Biclustering Algorithms}

This section provides a brief overview of biclustering algorithms embedded in BIDEAL. 

Cheng and Church (CC)\mbox{\cite{ref:CC}} proposed an algorithm to process expression data on the basis of Mean Squared Residue (MSR) score as
\begin{equation}
	\text{MSR}= \dfrac{1}{|I||J|}\sum_{i \in I, j \in J}(a_{ij}-a_{iJ}-a_{Ij}+a_{IJ})^2
    \label{eq:MSR score of CC}
\end{equation}

MSR measures coherency of genes and conditions using mean values and extract $\delta$-biclusters. Another effective algorithm FLexible Overlapped biClustering (FLOC) \cite{ref:Floc} was proposed. It performs probabilistic steps and find overlapped biclusters further refined using MSR score to overcome the effect of missing values in biclusters. The missing values often create random disturbances which affect the quality and slow down the operation of biclusters identification. The biclusters acquired by FLOC give better results for a larger matrix with smaller MSR in comparison to CC.

Dhillon \cite{ref:BSGP} used Bipartite Spectral Graph Partitioning (BSGP) to model data matrix as $G$=$(R,C,E)$. It is based on an exhaustive bicluster enumeration approach, which tries to find partitions of the minimum cut vertex in a bipartite graph between rows and columns. Considering the time and memory, it is quite expensive. BSGP approach can be represented as
\begin{equation}
  \begin{aligned}
  \resizebox{0.31\textheight}{!}{$
  cut(R_1 \cup C_1,...R_k \cup C_k)= \min\limits_{O_1,...O_k} cut(O_1,O_2...O_k)
 $}
 \end{aligned}
 \label{eq:BSGP}
\end{equation}

Order Preserving Sub-Matrices (OPSM) \cite{ref:OPSM} algorithm finds matrices, which have expression level in strictly increasing linear order. The algorithm uses a heuristic approach for biclustering. A sub-matrix can be said to be order preserving, if under the permutation of the conditions, the value of the gene expression data is linearly increasing or decreasing. 

Another approach proposed by Bergmann \textit{et al.} i.e. Iterative Search Algorithm (ISA) based on coherently overlapped biclusters, also referred as Transcription Modules (TM), can extract biclusters by iterative search from the gene expression data matrix \cite{ref:isa}.

In \cite{ref:kspectral}, Kluger \textit{et al.} proposed a spectral technique known as kSpectral to find biclusters based on Eigenvectors of the data matrix. Firstly, the datasets are normalized and then a singular value decomposition technique is applied on the micro array, where the constant part wise Eigenvalues give the checkerboard patterns in the sub-matrix. Finally, $k$-means clustering is applied to obtain the checkerboard structures from the data matrix.

In \cite{ref:ITL}, the authors presented the information-theoretic (ITL) formulation for biclustering. In this formulation, an optimization approach has been followed where the number of rows and column clusters are constraints and the task is to maximize the mutual information between clustered random variables. It can reduce the problem of high dimensionality and sparsity.

Murali \textit{et al.} \cite{ref:xMotif} proposed a representation for gene expression data called as conserved gene expression motifs or xMotifs. It tries to find largely conserved gene expression motifs from the given discretized data matrix. It uses a greedy approach that conserves row. A sub-matrix is said to be a conserved motif if the expression level of a gene is found consistent in the respective sub-matrix. Comparing distinct gene motifs for distinct conditions, we get to know of genes which are conserved in multiple conditions but are the in dissimilar state in various conditions.

In Plaid \cite{ref:plaid}, a bicluster is assumed to follow the statistical model and the binary least squares is used to fit the bicluster membership parameters. In this model, data matrix can be considered as a superposition of layers, where layer is a subset of genes and conditions of the data matrix. The data tries to fit in a plaid model can be expressed as
\begin{equation}
	a_{ij}= \sum_{k=1}^{B_{num}}{\theta_{ijk}+\rho_{ik}+\kappa_{jk}}
	\label{eq:Plaid}
\end{equation}

Binary Inclusion Maximal (BiMax) is based on fast divide and conquer approach \cite{ref:BiMax}. It tries to find all the bi-maximal biclusters which contains only one element. The algorithm requires discretization of the gene expression level matrix into a binary matrix by deciding a threshold.

Large Average Sub-Matrix (LAS) \cite{ref:LAS} is a statistically advanced algorithm which uses a Gaussian null model for gene expression data. It finds the bicluster to give the largest significance score which is defined as
\begin{equation}
    a_{ij} = \sum_{ k=1 }^{ B_{num} }l_r F \big(i \in 
    I_r, j \in J_r \big) + \xi
    \label{eq:LAS}
\end{equation}
The elements of the data matrix are subtracted from the mean of the significance score \eqref{eq:LAS} to form a residual matrix. The search is iteratively repeated until optimal $\varphi(\textbf{D})$ value falls below the predefined threshold.
 
Hochreiter \textit{et al.} \cite{ref:FABIA} presented a  multiplicative model biclustering algorithm i.e. Factor Analysis for Bicluster Acquisition (FABIA) that takes linear alliance of genes and conditions into account. In this model, the row and column vectors need to be multiple of each other. FABIA models the data matrix as the addition of $k$ biclusters and an additive noise. Here, the linear dependency of subsets of rows and columns can be described by outer product $\textbf{u} \times \textbf{v}^T$. The overall model is given by
\begin{equation}
	A = \sum_{ t=1 }^{ B_{num} } \textbf{u}_t \textbf{v}_t^T + \xi
	\label{eq: FABIA}
\end{equation}

In \cite{ref:BitBit}, bit-patterns are extracted from the data matrix using two phase process known as BitBit algorithm. The first phase includes a novel encoding process to divide the columns of the data matrix to a certain length determined by the minimum number of columns. In the second phase, biclustering of bit patterns takes place using selective search. Each pair of row generates a pattern. In BitBit, the comparison between rows takes place at bit level. To tackle excessive computation, iterative approach is used instead of divide and conquer approach as in BiMax by avoiding recursion and also additional traversals of the matrix a.k.a. BiSim \cite{ref:BiSim}.

Wang \textit{et al.} \cite{ref:MSVD} proposed Modular Singular Value Decomposition Multi-Objective Evolutionary biclustering (MSVD) algorithm. MSVD splits the gene expression data matrix into a set of sub-matrices with equal dimensions into a non-overlapping manner. Then, it projects the data obtained for the desired number of eigenvalues and applies $k$-means clustering to cluster them. 

Another algorithm QUalitative BIClustering (QUBIC) \cite{ref:QUBIC} based on graph theory approach is also embedded in BIDEAL. In QUBIC, the expression level of genes is expressed in a qualitative or semi-qualitative manner under multiple conditions as an integer value.

Tchagang \textit{et al.} proposed ROBA \cite{ref:ROBA}, where basic linear algebra techniques were used. There are three main steps in this algorithm. The first step involves preprocessing of data to handle missing values and noise. The second step decomposes given data matrix into binary matrices. The last step involves identification based on the type of bicluster. 

\section{BIDEAL: Accessible Validation Indices for Performance Measures}
\label{sec:Validation Indices}
Various validation indices as performance measures are used to check the quality of biclusters as described in further subsections.

\subsection{Jaccard Index}
\label{subsec:Jaccard Index}
Jaccard index \cite{ref:Fili} compares the biclusters obtained by applying the two biclustering algorithms and finding out the number of similar biclusters between them. Jaccard index gives a value of 0 if biclusters are dissimilar else 1. Jaccard index is defined as
\begin{equation}
	jac_c \big( \textbf{B}_1, \textbf{B}_2 \big) = \dfrac{ jac \big( \textbf{B}_1, \textbf{B}_2 \big) }{ \max \big( jac_c \big) }
	\label{eq:Jac}
\end{equation}

\subsection{SB Score}
\label{subsec:SB Score (SB Score)}
Differential co-expression ranking score a.k.a. SB score was proposed in \cite{ref:SB}. Considering two biclusters $\textbf{B}_1$ and $\textbf{B}_2$, where $\textbf{B}_1$ is formed by gene under the first set of conditions and $\textbf{B}_2$ is formed by the same gene with a second set of conditions. Chia \textit{et al.} proposed an algorithm to compare the goodness of gene w.r.t. two nonidentical set of conditions. If $\textbf{B}_1$ is good gene than there will be co-expression between gene and first set of conditions while differential co-expression between gene and second set of condition. The differential co-expression of $\textbf{B}_1$ can be measured as
\begin{equation}
 \resizebox{0.31\textheight}{!}{$
	SB(\textbf{B}_1) = \log \bigg( \dfrac{ \max(T_1(\textbf{B}_1) + \omega), \max(Q_1(\textbf{B}_1) + \omega) }{ \max(T_2(\textbf{B}_1) + \omega), \max(Q_2(\textbf{B}_1) + \omega)} \bigg)
    $}
    \label{eq: SB}
\end{equation}
where $\omega$ is used to offset the large ratios.

\subsection{Constant Variance}
\label{subsec:Constant Variance}
In \cite{ref:NKOPtimal}, corresponding variance of genes/ conditions is taken into consideration where the variance is the average of the sum of Euclidean distances between rows and columns of bicluster. Higher the value of the variance, lower the quality of the bicluster. The expression of the variance is given by
\begin{equation}
	\text{var} = \sum_{i \in I, j \in J} ( a_{ij} - a_{IJ} )^2
	\label{eq:Constant Variance}
\end{equation}

\subsection{Sign Variance}
\label{subsec:Sign variance}
For a coherent bicluster, the value of sign variance is lower \cite{ref:MTBA}. It is same as constant variance except it preprocesses the data matrix into sign matrix and then estimates variance.

\subsection{Hausdorff Distance}
\label{subsec:Hausdorff Distance}
The Hausdorff distance \cite{ref:NKBIBM} calculates the distance between the pair of sub-matrices obtained from the gene expression data matrix. It is maximum for traversal from the element of first bicluster to the nearest element of second bicluster and signifies dissimilarity. Mathematically, it can be written as
\begin{equation}
  \begin{aligned}
  HD \big( \textbf{B}_1, \textbf{B}_2 \big) = \max \big \{ \text{sup}_{ b_1 \in \textbf{B}_1} \text{inf}_{ b_2 \in \textbf{B}_2 } d(b_1,b_2),\\ \text{sup}_{ b_2 \in \textbf{B}_2 } \text{inf}_{ b_1 \in \textbf{B}_1 } d(b_1,b_2) \big \}
  \end{aligned}
  \label{eq:HD}
\end{equation}
 
\subsection{Mean Squared Residue}
\label{subsec:Mean Residue Score}
To calculate mean squared residue, the mean square error (MSE) of each bicluster is calculated \cite{ref:CC}. Then overall MSE can be calculated by taking the mean of individual values.

\section{BIDEAL: Key Features and GUI}
\label{sec:BIDEAL: Key Features and GUI}
BIDEAL integrates various biclustering algorithms into a stand-alone application of graphical user interface (GUI) developed using MATLAB. It is executable on Windows as well as on Linux operating system. BIDEAL includes several functions to preprocess the raw data, validate, and visualize the biclusters. The key features of BIDEAL are as follows:
\begin{figure*}
\centering
\includegraphics[width=180mm]{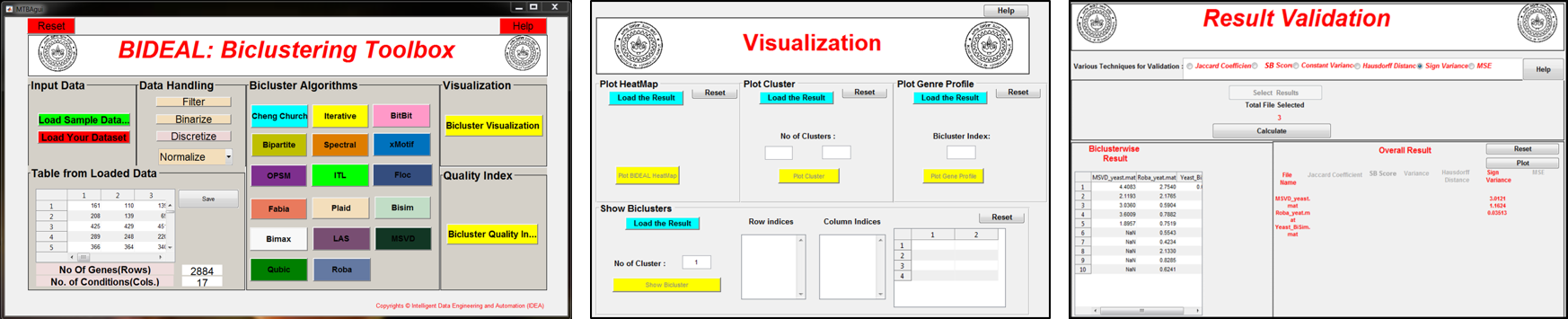}
\caption{Graphical User Interface of BIDEAL. \textbf{From left to right:} Homepage, Visualization page, and Validation page.}
\label{fig:BIDEAL GUI}
\end{figure*}

\subsection{Data Preprocessing}
\label{subsec:Preprocessing}
BIDEAL includes four preprocessing methods, i.e. filtering, binarization, discretization, and normalization. Filtering is used to eliminate the effect of Not a Number (NaN) spots and missing values from the data. Binarization is used to convert a numerical feature vector into a Boolean, it is mostly useful for downstream probabilistic estimators which assume that the input data is distributed according to a multi-variate Bernoulli distribution. Discretization, a.k.a. quantization/ binning, is used to transform continuous features into discrete values. Some specific datasets with continuous features may not be linearly correlated with the target and are not able to handle with feature selection methods. In such cases, obtaining an interpretable explanation of such features won’t be easy. However, this type of data may be benefited from discretization because it can transform the dataset of continuous attributes to one with only nominal attributes. Normalization is used for scaling the individual samples to have unit norm. In general min-max and z-score normalization are used when data come from the normal distribution. However, biomedical data or most of the clinical research data do not follow the normal distribution because they are mostly skewed. For this purpose, logarithmic transformation bistochastization and item independent re-scaling of rows and columns are used. The log transformation decreases the variability of data and bistochastization makes all rows and columns to have the same mean value and the matrix is repeatedly normalized until convergence, whereas, in the independent row and column normalization of rows sum to a constant and columns sum to a distinct constant \cite{ref:kspectral}.
 
\subsection{Largest Number of Biclustering Algorithms}
\label{subsec:Biclustering Algorithms}
For biclusters generation, $17$ biclustering algorithms have been embedded in BIDEAL that is maximum among all the available toolbox listed in Table \mbox{\ref{table:Comparison of features comprised with different biclustering toolboxes}}. It provides flexibility to select biclustering algorithms according to the nature of data. Availability of all algorithms at a single platform allows to analyze the data with minimal efforts.

\subsection{Initial Parameter Setting of Algorithms} 
\label{subsec:parameter}
Without a prior knowledge of algorithms, the parameters setting is quite challenging for naive user. BIDEAL facilitates the initial value of parameters as provided in the original published work which users can easily change if needed.

\subsection{Robust Bicluster Generation}
\label{subsec:robust}
BIDEAL offers several ways to ensure a smooth and robust bicluster generation. For example, the filtering option is availed to reduce the effect of NaN and missing values present in the dataset.

\subsection{Identification of Cluster Type}
\label{subsec:coherent}
BIDEAL offers validation indices to determine the type of biclusters. For example, the constant variance can identify constant bicluster, whereas sign variance allows to identify bicluster where coherent sign changes on rows and columns.

\subsection{Similarity Measures}
\label{subsec:correlation}
BIDEAL offers two validation indices, i.e. Jaccard index and Hausdorff distance to measure the similarity and dissimilarity, respectively between two biclusters. The value of Jaccard index of a particular biclustering algorithm varies from $0$ to $1$ depending upon the level of similarity. Hausdorff distance, widely used in several applications, can also measure the distance between two distinct biclusters. For example, in Yeast dataset, Jaccard index values were calculated for CC algorithm and it can be seen that results obtained from other algorithms were dissimilar from CC as Jaccard index values were very less for all other algorithms.

\subsection{User-Friendly Interface}
\label{subsec:userfriend}
BIDEAL offers a user friendly GUI which is easy to use for bicluster analysis including generation, visualization, and validation. The unique features of this interface are:
\begin{enumerate}[label=(\alph*)]
    \item BIDEAL is a self contained concise toolbox with all the relevant information present in it. It provides immediate visual results and effect of each action.
    \item In many cases, the installation of toolbox depends on other components like language, which in general is not availed with toolbox package. To ease the installation, the stand-alone executable files are packaged with MATLAB run-time compiler in BIDEAL. This enables the user to just click and install the ready to use biclustering algorithms.
\end{enumerate}

\subsection{Implementation and GUI}
\label{subsec:GUI}
BIDEAL has been developed using MATLAB which integrates various features into a stand-alone application. The GUI of developed BIDEAL toolbox comprises of the following steps for biclusters generation, validation, and visualization:
\begin{enumerate}[label=(\roman*)]
    \item The home page of BIDEAL is shown in Fig. \ref{fig:BIDEAL GUI}. At first, the dataset should be loaded. It can be either a sample or user-defined dataset.
    \item The data can be preprocessed using filtering, binarization, normalization, or discretization.
    \item Select the required algorithm to generate biclusters. User will be prompted to feed input parameters else BIDEAL will consider the default values.
    \item Generated results can be saved in .mat file.
    \item Click the \textit{Bicluster Visualization} button on the home page to visualize the biclusters. Any of the available three options on visualization page i.e. heat map, cluster plot, or gene profile can be clicked to visualize the result.
    \item Click the \textit{Bicluster Quality Index} button to access the validation indices. The validation page displays individual bicluster or overall biclusters result.
    \item Press \textit{Reset} button to again access the home page.
\end{enumerate}

\begin{figure*}
    \centering
    \includegraphics[width=165mm]{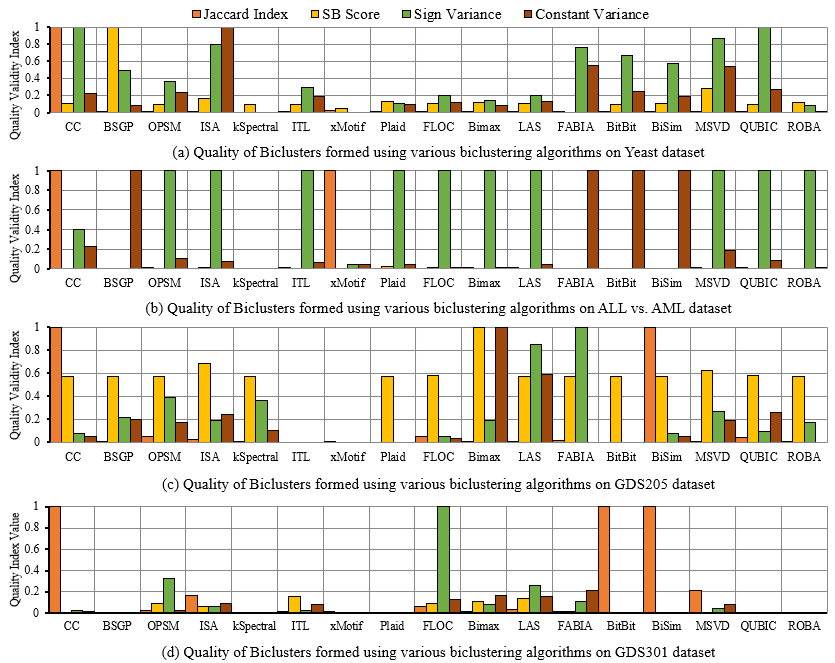}
    \caption{Validation indices on various datasets.}
    \label{fig:Validation Indices on various datasets}
\end{figure*}
\begin{table*}
  \caption{Number of biclusters obtained with biclustering algorithms available in BIDEAL on four datasets}
  \label{table:Number of Biclusters using 17 Algorithms Available in BIDEAL on Four Datasets}
  \centering
  \resizebox{.97\textwidth}{!}{%
  \begin{tabular}{|c|c|c|c|c|c|c|c|c|c|c|c|c|c|c|c|c|c|c|}
  \hline
  \backslashbox{\textbf{Datasets}}{\textbf{Algorithms}}
  & \makecell{\textbf{CC}\\\cite{ref:CC}} & \makecell{\textbf{BSGP}\\\cite{ref:BSGP}} & \makecell{\textbf{OPSM}\\\cite{ref:OPSM}} & \makecell{\textbf{ISA}\\\cite{ref:isa}} & 
  \makecell{\textbf{kSpectral}\\\cite{ref:kspectral}} & \makecell{\textbf{ITL}\\\cite{ref:ITL}} & \makecell{\textbf{xMotif}\\\cite{ref:xMotif}} & \makecell{\textbf{Plaid}\\\cite{ref:plaid}} & \makecell{\textbf{FLOC}\\\cite{ref:Floc}} & \makecell{\textbf{BiMax}\\\cite{ref:BiMax}} & \makecell{\textbf{LAS}\\\cite{ref:LAS}} & \makecell{\textbf{FABIA}\\\cite{ref:FABIA}} & \makecell{\textbf{BitBit}\\\cite{ref:BitBit}} & \makecell{\textbf{BiSim}\\\cite{ref:BiSim}} & \makecell{\textbf{MSVD}\\\cite{ref:MSVD}} & \makecell{\textbf{QUBIC}\\\cite{ref:QUBIC}} & \makecell{\textbf{ROBA}\\\cite{ref:ROBA}} \\
  \hline
  \textbf{Yeast}\cite{ref:CC} & 100 & 10 & 16 & 16 & 0 & 1 & 97 & 4 & 20 & 75 & 20 & 2 & 212 & 1547 & 13 & 10 & 10104 \\
  \hline
  \textbf{ALL vs. AML}\cite{ref:ALLvsAML} & 1 & 0 & 37 & 500 & 0 & 100 & 89 & 4 & 20 & 100 & 52 & 5 & 0 & 0 & 100 & 0 & 32591 \\
  \hline
  \textbf{GDS205}\cite{ref:GDS205} & 1 & 7 & 7 & 13 & 6 & 0 & 5 & 0 & 20 & 11 & 5 & 5 & 0 & 1 & 3 & 10 & 3925 \\
  \hline
  \textbf{GDS301}\cite{ref:GDS301} & 1 & 0 & 10 & 0 & 0 & 100 & 39 & 0 & 20 & 100 & 5 & 4 & 1 & 1 & 1 & 0 & 0 \\
  \hline
\end{tabular}
}
\end{table*}

\section{BIDEAL: Testing and Validations on Benchmark Datasets}
\label{sec:Case Studies}
\begin{figure}
\centering
\includegraphics[width=88mm]{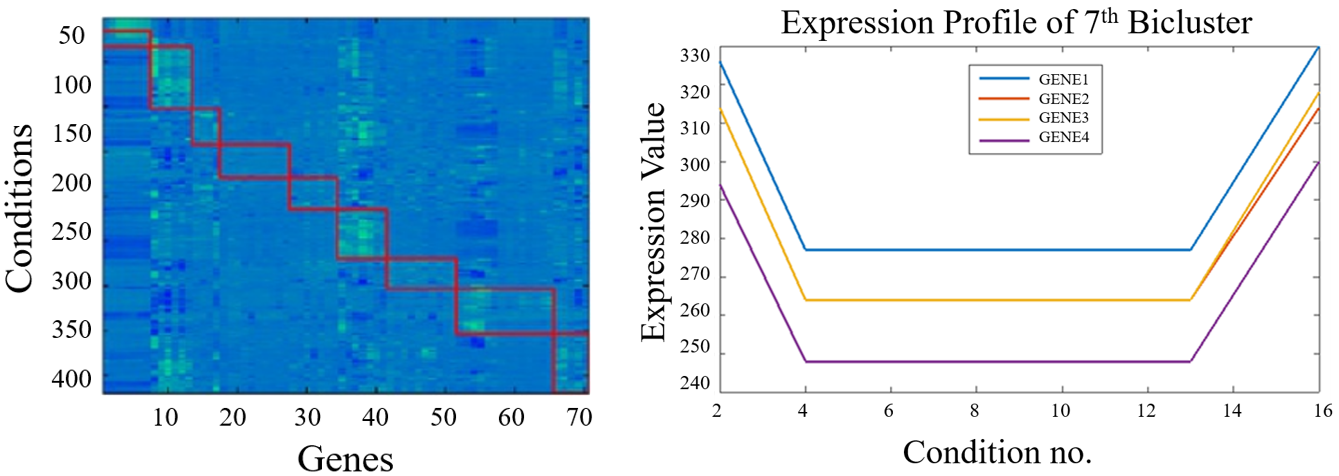}
\caption{\textbf{Left:} Heat map plot and \textbf{Right:} Gene plot using CC algorithm for generated bicluster on Yeast dataset.}
\label{fig: heat map 1}
\end{figure}
To demonstrate the utility of gene expression profiling by generation of patterns or biclusters through a single platform decreases user efforts. Hence, BIDEAL offers a user friendly interface to decrease the cumbersomeness faced during  the biclusters formation. In this section, the experiments and validation on four benchmark datasets have been provided using BIDEAL. The four datasets used are Saccharomyces cerevisiae cell cycle dataset (Yeast) \cite{ref:CC} with $2,884$ genes and $17$ conditions, Leukemia (ALL vs. AML) dataset \cite{ref:ALLvsAML} with $3,571$ genes and $72$ conditions, Mammary tissue profile dataset (GDS205) \cite{ref:GDS205} with $822$ genes and $8$ conditions, and Ligand screen in B cells dataset (GDS301): Epstein Barr virus-induced molecule-1 \cite{ref:GDS301} with $16,271$ genes and $11$ conditions. The biclusters formed on these four benchmark datasets are further validated using validation indices available in BIDEAL as depicted in  Fig. \ref{fig:Validation Indices on various datasets}. Table \ref{table:Number of Biclusters using 17 Algorithms Available in BIDEAL on Four Datasets} tabulates the number of biclusters obtained using $17$ biclustering algorithms embedded in  BIDEAL. Since Yeast \cite{ref:CC} and ALL vs. AML \cite{ref:ALLvsAML} datasets are preprocessed therefore GDS205 and GDS301 were preprocessed before execution of the biclustering algorithms. In further subsections, the findings of BIDEAL have been discussed in detail.

\subsection{Saccharomyces Cerevisiae Cell Cycle (Yeast) Dataset}
\label{subsec:Biological Dataset (Yeast)}
Yeast dataset \cite{ref:CC} comprises of $2,884$ genes and $17$ conditions. The objective of this dataset is the identification of genes whose mRNA levels are regulated by the cell cycle. The number of biclusters generated on Yeast dataset using BIDEAL have been reported in Table \ref{table:Number of Biclusters using 17 Algorithms Available in BIDEAL on Four Datasets}. The table depicts that among all algorithms ROBA generates highest number of biclusters whereas kSpectral fails to produce any bicluster i.e. 0. It is due the fact that kSpectral did not find any distinctive checkerboard patterns in Yeast dataset. On the other hand, ROBA utilizes simple linear algebraic methods instead of complex optimization and extracted highest i.e. 10104 number of biclusters. Since the hierarchy of biclustering algorithms is application specific therefore one cannot measured their utility in terms of number of bicluster like BiSim forms $1547$ biclusters whereas ITL and FABIA extracted only 1 and 2 biclusters respectively. However all of them have their own biological significance. CC forms $100$ biclusters which cover approximately $97\%$ genes and approx. $82\%$ of conditions. Fig. \ref{fig: heat map 1} shows a sample heat map and gene plot using CC algorithm for generated bicluster on Yeast dataset. 
BSGP and QUBIC reported $10$ biclusters, whereas FABIA and Plaid had very few biclusters with fewer genes and conditions. BitBit gave $212$ biclusters while kSpectral failed to produce any bicluster which signifies that this model do not fit with the given dataset. OPSM and ISA reported the same number of biclusters. Considering the quality of obtained biclusters, it was noted that the biclusters obtained using BiSim, FABIA, and kSpectral had no similarity w.r.t. CC in the context of Jaccard index. On the other hand, ITL, Plaid, BitBit, and ISA had very low similarity. BSGP and MSVD gave higher similarity while ROBA had the maximum similarity among all. According to sign variance metric, the biclusters obtained using CC, Plaid, ISA, and FABIA were less coherent while ROBA, BSGP, and BiMax gave strong coherent biclusters. LAS, BiSim, and MSVD were giving average coherent biclusters. While measuring the quality of biclusters using constant variance, it was inferred that BSGP, MSVD, BiMax formed better biclusters while ISA and Plaid gave higher values of constant variance indicating lower quality of biclusters. LAS, CC, BitBit, ITL, and FLOC gave an average type of biclusters.

\subsection{Leukemia (ALL vs. AML) Dataset}
\label{subsec:Leukemia dataset (ALL vs. AML)}
Leukemia dataset comprises of two subtypes of leukemia cancer i.e. Acute Myeloid Leukemia (AML) and Acute Lymphoblastic Leukemia (ALL). It has $3571$ genes and $72$ conditions. For ALL vs. AML dataset, also ROBA reported highest number of biclusters $32,591$ biclusters due to its ability to extract more than one type of biclusters in given dataset. As mentioned earlier various biclustering algorithms are able extract specific patterns from dataset. For ex. BSGP works better when dataset can be modelled using bipartite graph efficiently whereas kSpectral is well known to extract checkerboard patterns in data. In case of this dataset both patterns were not applicable therefore 0 biclusters were reported. On other hand BitBit and BiSim are known to search patterns in less time by traversing the binarized data matrix with tuned parameters. As shown in Table \ref{table:Number of Biclusters using 17 Algorithms Available in BIDEAL on Four Datasets} BSGP, kSpectral, BitBit and BiSim failed to produce any bicluster. BiMax successfully extracts 100 inclusive maximal biclusters from this dataset. It is interesting to notice that ITL, BiMax and MSVD produced same number of biclusters i.e. 100 though their objective functions are different from each other. ITL tries to preserve mutual info whereas BiMax follows divide and conquer strategy and MSVD is inspired from linear algebra technique.  CC formed only one bicluster which has all genes and conditions. LAS, OPSM and xMotif resulted 52, 37 and 89 bicluster respectively. FABIA and Plaid extracted only  5 and 4  biclusters due to presence of less conditions and few layers as per plaid model. Considering the Jaccard index similarity, xMotif and CC values were high. CC and xMotif had a negative score which indicates differential co-expression. According to sign variance, CC gave coherent biclusters as it had the lowest value while high value of FLOC and BiMax indicates less coherent cluster. Rest of the algorithms generated biclusters with average coherency. From constant variance values, it can be inferred that ISA gave very low quality biclusters.

\subsection{Mammary Tissue Profile (GDS205) Dataset}
\label{subsec:Mammary tissue profile dataset (GDS205)}
GDS205 \cite{ref:GDS205} comprises of $822$ genes and $8$ conditions. For this dataset again ROBA resulted in high  number of biclusters i.e. 3925. This indicates there are overlapped gene and sample sets where genes are involved in several biological pathways. Rest of the biclustering algorithms, embedded in BIDEAL extracted approximately 12 biclusters only. BiMax successfully extracted 11 subsets of genes and conditions whereas BiSim only extracted 1 bicluster. FABIA extracted only 5 biclusters which signifies that GDS205 dataset is not influenced by heavy-tailed distribution. For this dataset use of FLOC algorithm over the CC is clearly shown. FLOC resulted in $20$ biclusters without being effected by random interference whereas as CC produced only $1$ bicluster. BSGP and OPSM both gave $7$  biclusters indicating presence of order-preserving sub-matrices in GDS201. kSpectral and xMotif resulted in $6$ and $5$, respectively. LAS and MSVD discovered $5$, $3$ biclusters, respectively. Qubic identified $10$ checkerboard pattern present in data. For this dataset ITL, Plaid, and BitBit failed to provide any bicluster. Plaid did not find any shift biclusters in this dataset whereas ITL fails to find co entropy based subsets genes and conditions.
Now considering the validity of these bicluster we found that in terms of sign variance, CC and QUBIC resulted in very low value i.e. more coherent biclusters but biclusters produced by LAS were not coherent hence it had high value of sign variance. According to the constant variance, CC and QUBIC produced best biclusters, but FLOC gave the high value of constant variance, which meant that the quality of the biclusters was not good. Jaccard indices were calculated w.r.t. CC like others. It interprets that the biclusters formed by BSGP and MSVD had the lowest similarity with the biclusters formed by CC. It can also be concluded that CC and QUBIC produced better biclusters for this dataset.

\subsection{Ligand Screen in B Cells (GDS301) Dataset}
\label{subsec:Ligand screen in B Cells (GDS301)}
GDS301 dataset comprises of $16,271$ genes and 11 conditions collected by culturing B Cells with Ligand to perform temporal analysis. As shown in Table \ref{table:Number of Biclusters using 17 Algorithms Available in BIDEAL on Four Datasets} BiMax produced maximum number of biclusters i.e. $100$. This signifies 100 biclusters were found with values of  1s by enumeration. ITL also discovered same number of biclusters by extracting mutual information between genes and conditions.  BSGP, kspectral, and Plaid failed to produce any bicluster. Plaid discovers interesting pattern with multivariate data whereas kSpectral identifies biclusters only if genes are co-regulated with expression levels. FABIA reported to extract 4 biclusters. CC, ISA, BitBit, BiSim, all reported one bicluster having all genes and conditions in that bicluster. This means algorithms failed to extract the patterns from dataset. Though MSVD formed one bicluster where all conditions were present but only $3,257$ genes were matched.  In terms of Jaccard index, BitBit and BiSim had maximum similarity with CC, whereas ITL and BiMax had less similarity with CC. In terms of sign variance, xMotif and CC gave coherent biclusters but biclusters formed by FLOC were not coherent enough. Constant variance values were mostly similar i.e. FABIA produced maximum constant variance among all.

\subsection{Biological Significance}
\label{subsec:Biological Significance}
The biological significance of biclustering algorithms refers to the identification of subset of genes clustered with similar subset of conditions to form a pattern or bicluster. The biclusters are useful for disease identification, biomarkers generation, gene-drug association, etc. The reliability of these biclusters are justified using various evaluation measures. BIDEAL provides constant variance and sign variance as evaluation  measures to check the coherency, significance, and reliability of biclusters obtained using various biclustering algorithms . In terms of coherency, for Yeast dataset, biclusters generated using FLOC, Bimax, LAS, and ITL algorithms had low sign variance and constant variance. In ALL vs. AML dataset, most of the algorithms failed to generate significantly coherent biclusters except CC and xMotif algorithms. In GDS205 dataset, CC and BiSim algorithms produced coherent biclusters whereas, in GDS301 dataset, CC, ITL, and ISA algorithms produced coherent biclusters. Another evaluation measure, i.e. SB score, is also embedded in BIDEAL. The SB score was quite low for Yeast dataset except for the biclusters generated using BSGP algorithm. It shows that the obtained biclusters had more co-expression level for two conditions among genes. In ALL vs. AML dataset, generated biclusters have differential co-expression among genes and conditions because the value of SB score was almost absent. GDS205 dataset reported the high value of SB score which signifies the more co-expression ranking among genes w.r.t. two sets of conditions. In each dataset, at least one algorithm had reported similar bicluster as CC algorithm, for example ITL in case of ALL vs. AML dataset, whereas BiSim in GDS205 dataset. As presented in Table \ref{table:Number of Biclusters using 17 Algorithms Available in BIDEAL on Four Datasets}, it can be seen that for $3$ datasets, ROBA gave an exceptionally large number of biclusters which means overlapping biclusters were generated, FABIA and plaid resulted in less number of biclusters for all datasets, FLOC generated a constant number of biclusters i.e. $20$. For GDS301, only CC, OPSM, ITL, xMotif, FLOC, BiMax, LAS, ISA, MSVD, and FABIA had some result and BiSim and BitBit were quite similar to CC. In case of Yeast dataset, kSpectral failed to produce any bicluster while ITL, Plaid, and BitBit gave no bicluster on GDS205 dataset. Most of the biclusters formed using xMotif, BiSim, QUBIC, BSGP, and CC are of $\mu$-type which indicates clusters with strong instance and attribute effect. MSVD, FLOC, ISA, and BiMax generated biclusters are of T-type hence these biclusters are with strong instance effect.

\subsection{Execution Time and Size of Dataset}
\label{subsec: Execution Time and Size of Dataset}
The proposed toolbox integrates various biclustering algorithms on a single platform therefore to measure the execution time one needs to note the execution time of each algorithm. Since the complexity of the biclustering problem relies on the dataset and the objective function therefore its execution time can vary from few seconds to hours. For example on the Yeast dataset, CC, xMotif, and BiMax takes less than 5 seconds to compute biclusters; BSGP, ISA, kSpectral, and FLOC take around 1 minutes to compute biclusters; BitBit and QUBIC extracts biclusters in 30 minutes; and BiSim executes in 90 minutes. Moreover, considering the maximum file sizes can be handled, the proposed toolbox has been validated for the dataset with maximum size of 25 MB. The test has been performed on Yeast dataset of file size 198KB, ALL vs. AML dataset of file size 656KB, GS205 dataset of file size 120KB, and GDS301 dataset of file size 25 MB.

\section{Conclusions}
\label{sec:Conclusions}
The proposed \enquote{BIDEAL} toolbox in this paper has been developed to generate, validate, and visualize the biclusters from any data on a single platform. It integrates $17$ famous biclustering algorithms, $6$ validation indices, and $3$ visualization methods for comprehensive data interpretations. Additionally, it provides preprocessing module to remove outliers and NaN spots from the data which helps to rectify issues related to null values, discrete matrix, etc. The proposed toolbox has been tested and validated on four benchmark gene expression datasets i.e. Yeast, ALL vs. AML, GDS205, and  GDS301. It was inferred that each algorithm of BIDEAL can generate distinct set of biclusters from the same data; therefore, the selection of appropriate technique is required. The diverse nature of BIDEAL with various validation indices and visualization methods has been proven effective for selection of best biclusters. Information retrieval from data mainly depends on the type of local patterns, whether it has overlapping and constant biclusters, or noisy data. We hope that the availability of BIDEAL will help the research community by widespread use of biclustering algorithms to identify coherent groups in data which is very useful in disease subtype identification. Furthermore, the toolbox can help to cater the data analysis needs, and it is being offered free to the community.

\vfill
\begin{IEEEbiography}
[{\includegraphics[width=1in,height=1.65in,clip,keepaspectratio]{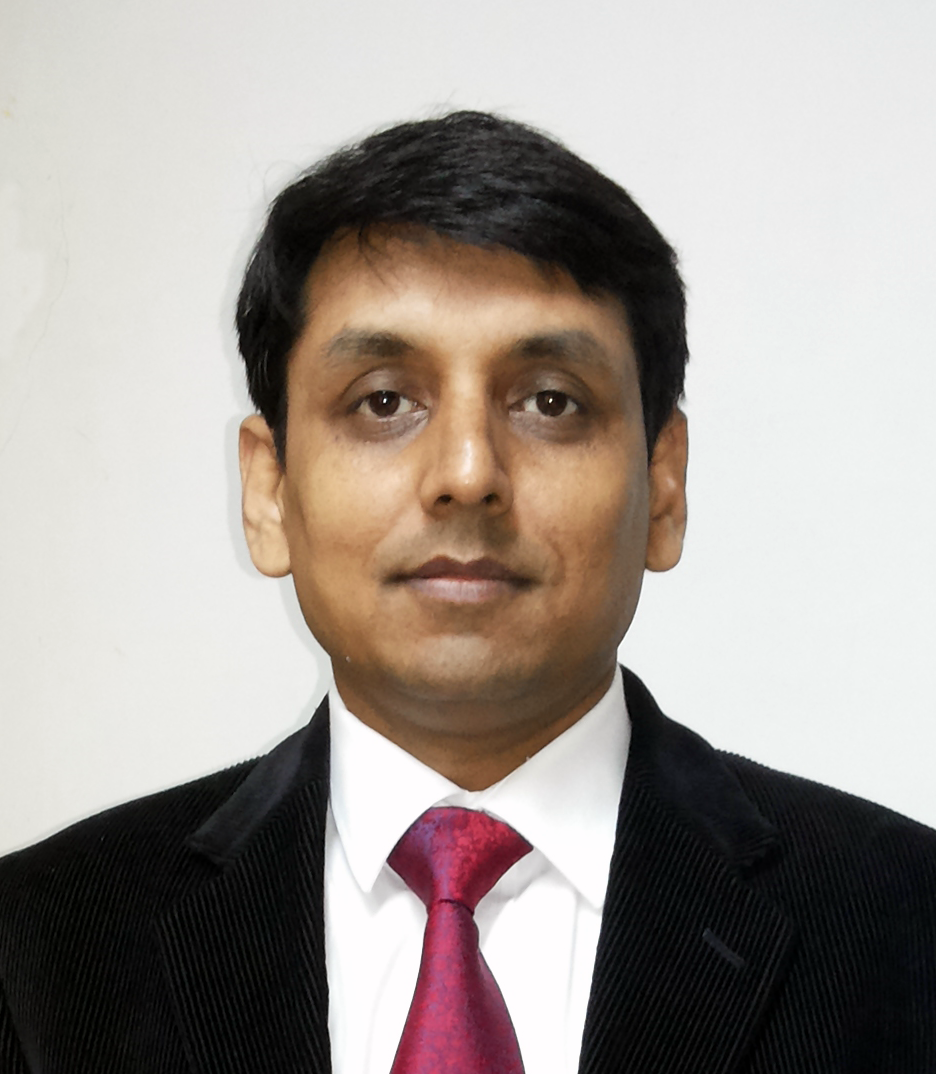}}]%
{Nishchal K. Verma}
(SM'13) is a Professor in Department of Electrical Engineering and Inter-disciplinary Program in Cognitive Science at Indian Institute of Technology Kanpur, India. He obtained his PhD in Electrical Engineering from Indian Institute of Technology Delhi, India. He is an awardee of Devendra Shukla Young Faculty Research Fellowship by Indian Institute of Technology Kanpur, India for year 2013-16.
His research interests include big data analysis, deep learning of neural and fuzzy networks, machine learning algorithms, computational intelligence, computer vision, brain computer/machine interface, intelligent informatics, soft-computing in modelling and control, internet of things/ cyber physical systems, cognitive science and intelligent fault diagnosis systems, prognosis and health management. He has authored more than 200 research papers.\\
Dr. Verma is an IETE Fellow. He is currently serving as a Guest Editor of the IEEE Access special section \enquote{Advance in Prognostics and System Health Management}, an Editor of the IETE Technical Review Journal, an Associate Editor of the IEEE Transactions on Neural Networks and Learning Systems, an Associate Editor of the IEEE Computational Intelligence Magazine, an Associate Editor of the Transactions of the Institute of Measurement and Control, U.K. and Editorial Board Member for several journals and conferences.
\end{IEEEbiography}

\begin{IEEEbiography}[{\includegraphics[width=1in,keepaspectratio]{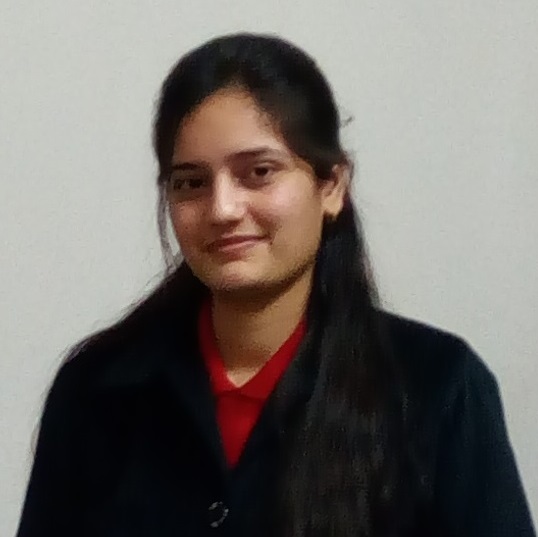}}]{Teena Sharma}
received her B.Tech. degree in Electronics and Communication Engineering in 2013 from UPTU, Lucknow, India. She has completed her M.tech in 2014 from Banasthali University, Rajasthan, India. Currently, She is a PhD Scholar in Dept. of Electrical Engineering at Indian Institute of Technology Kanpur, India. Her Research interests fall under Machine Learning, Computer Vision, and its applications.
\end{IEEEbiography}

\begin{IEEEbiography} 
[{\includegraphics[width=1.0in,height=2.5in,clip,keepaspectratio]{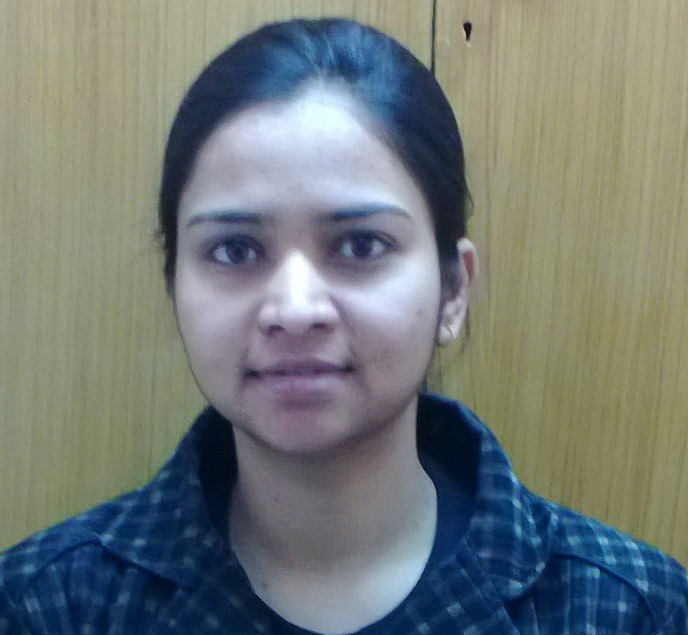}}]%
{Sonal Dixit}
received her B.E. from RGPV University in 2009, and Masters from Banasthali University in 2011. She is currently a doctoral student at IIT Kanpur. Her research  interests fall mainly in the field of condition based monitoring of rotary machines,  deep learning and its applications, computational intelligence, and natural language processing. 
\end{IEEEbiography}

\begin{IEEEbiography}
[{\includegraphics[width=1in,height=1.65in,clip,keepaspectratio]{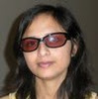}}]%
{Pooja Agrawal}
 received her Ph.D. degree in Aerospace Engineering at IISc Bangalore, India in 2017. She is currently working as a Post-Doctoral Fellow in the Dept. of Electrical Engineering, IIT Kanpur, India. Her research interests include theory and applications of computational intelligence, knowledge discovery, intelligent data mining algorithms and vision based control and guidance of UAVs. 
 \end{IEEEbiography}

 \begin{IEEEbiography}
 [{\includegraphics[width=1in,height=1.65in,clip,keepaspectratio]{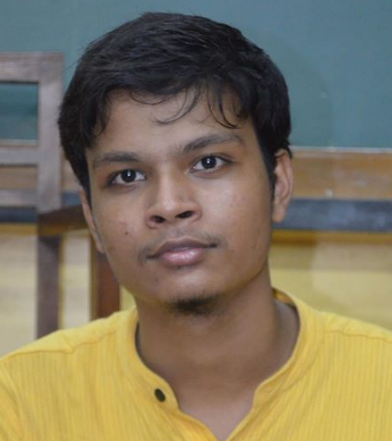}}]%
 {Sourya Sengupta} was born on 23rd September,1995 at Kolkata, India. Currently he is pursuing his Master's from University of Waterloo. He received B.E. degree in Electrical Engineering at Jadavpur University, Kolkata India. He passed Madhyamik and Higher Secondary Examination from Ramakrishna Mission Boys’ Home Higher Secondary School,Rahara on 2012 and 2014 respectively. He also qualified WBJEE, JEE Mains and JEE Advanced on 2014. His research interests fall mainly in the field of biomedical signal processing, bioinformatics, cognitive neuroscience .
\end{IEEEbiography}
\begin{IEEEbiography}
[{\includegraphics[width=1in,height=1.65in,clip,keepaspectratio]{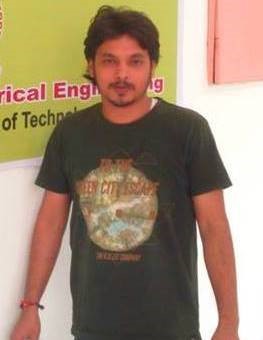}}]%
{Vikas Singh} is working toward the PhD degree
in the Department of Electrical Engineering,
Indian Institute of Technology Kanpur, India.
His research interests include machine learning,
deep learning, big data, intelligent data mining,
bioinformatics, and fuzzy systems and its
applications.
\end{IEEEbiography}

\end{document}